\documentclass[aps,pra,twocolumn,showpacs,superscriptaddress,floatfix]{revtex4}
\usepackage{epsfig}
\usepackage{amsmath}

\newcommand{\id}{{\sf 1 \hspace{-0.3ex} \rule{0.1ex}{1.52ex}\rule[-.01ex]{0.3ex}{0.1ex}}}
\newcommand{\ignore}[1]{}

\begin{document}

\title{Remarks on Duality Transformations and Generalized Stabilizer States}

\author{Martin B. Plenio}
\affiliation{Institute for Mathematical Sciences, Imperial College
London, 53 Prince's Gate Exhibition Road, London SW7 2PG, UK}
\affiliation{QOLS, Blackett Laboratory, Imperial College London, Prince Consort
Road, London SW7 2BW, UK}
\date{\today}

\begin{abstract}
We consider the transformation of Hamilton operators under various
sets of quantum operations acting simultaneously on all adjacent pairs of
particles. We find mappings between Hamilton operators analogous
to duality transformations as well as exact characterizations of ground
states employing non-Hermitean eigenvalue equations and use this
to motivate a generalization of the stabilizer formalism to
non-Hermitean operators. The resulting class of states is larger
than that of standard stabilizer states and allows for example for
continuous variation of local entropies rather than the discrete
values taken on stabilizer states and the exact description of certain ground
states of Hamilton operators.
\end{abstract}

 \maketitle

\section{Introduction}
The investigation of quantum many-body systems with the tools of
entanglement theory \cite{Plenio V 07} has recently received
considerable attention addressing long-standing questions in the
latter employing new methods developed in the former. This
includes for example the scaling of block entropies
\cite{blockentropy} and geometric entropies \cite{Plenio CDE 05},
a new improved understanding and generalization of numerical
methods such as DMRG on the basis of matrix-product states
\cite{Fannes NW 92,Verstraete C 04} and the development of novel
approaches based on new classes of efficiently describable quantum
states such as weighted graph states \cite{Anders PDVB 06,Dur
HHLPB 05}. Most of these methods are based on the efficient
description of quantum states and are taking place in the
Schr{\"o}dinger picture while the numerical algorithm described
in \cite{Anders PDVB 06} is a mixture of Schr{\"o}dinger and
Heisenberg picture,considering both transformations of Hamiltonians
and description of states. In this note consider only transformations
on Hamiltonians and operators characterizing states.

\section{Duality transformations}
In the following I will consider the effect of sequences of one-
and two-qubit quantum gates, such as control phase gates and
single qubit operations, in a spirit not dissimilar to the action
of a cellular automaton \cite{Schumacher W 04} on Hamilton
operators. 
Here and in the following we employ the notation
\begin{equation}
    X = \left(\begin{array}{cc} 0 & 1 \\ 1 & 0
    \end{array}\right), \;\;
    Y = \left(\begin{array}{cc} 0 & -i \\ i & 0
    \end{array}\right), \;\;
    Z = \left(\begin{array}{cc} 1 & 0 \\ 0 & -1
    \end{array}\right)
\end{equation}
for the Pauli-operators and $\id$ for the identity to simplify
notation.

\begin{figure}[th]\vspace*{-0.4cm}
\centerline{
\includegraphics[width=9cm]{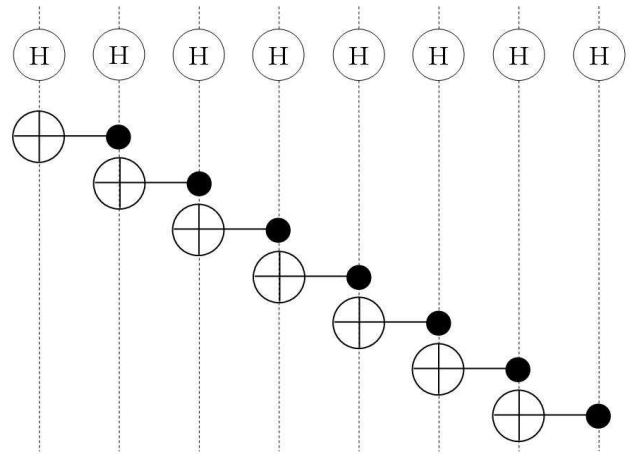}
}
\vspace{-.5cm} \caption{\label{fig2} Time progresses from bottom
to the top.  Sequence of application of CNOT gates to a set of $N$
qubits followed by the application of Hadamard gates implementing
the mapping $H|0\rangle = (|0\rangle+|1\rangle)/\sqrt{2}$ and
$H|1\rangle = (|0\rangle-|1\rangle)/\sqrt{2}$. }
\end{figure}

\begin{figure}[th]\vspace*{-0.4cm}
\centerline{
\includegraphics[width=8.5cm]{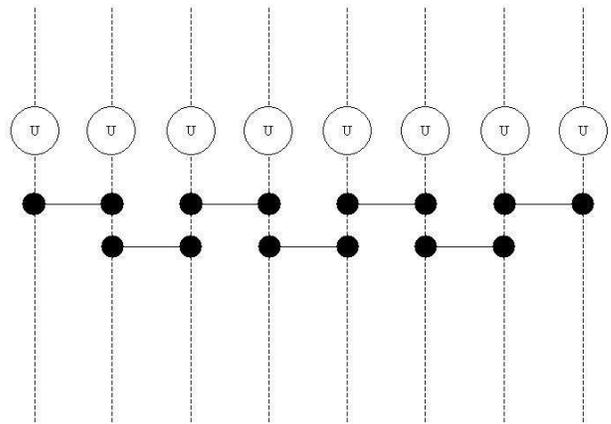}
}
\vspace{-.5cm} \caption{\label{fig1} Time is progressing from
bottom to the top. A cellular automaton first applies controlled
phase gates (they are symmetric with respect to interchange of 
control and target) between nearest neighbours and then single 
qubit gates
$U$ such that the combined action maps the Hamilton operator eq.
(\ref{zxz}) onto a single particle Hamiltonian that is diagonal in
the computational basis.}
\end{figure}

In general, it cannot be expected that with a few simple steps of
the nature presented in Fig. \ref{fig2} one can diagonalize a
Hamilton operator (even though later on in this section such an
example will be discussed which is then used to motivate the definition
of generalized stabilizer states). It may however be possible to
transform Hamilton operators into each another, thus establishing
useful equivalences between seemingly different systems. With this
aim in mind we would like to explore what effects the application
of sequences of quantum gates may have on Hamilton operators
describing quantum many body systems. To avoid problems with
operator ordering we will consider finite translation invariant
systems with open boundary conditions. This will allow me to use
settings such as those in Fig. \ref{fig2} instead of the standard
cellular automaton in the Margolus partitioning (see e.g. Fig.
\ref{fig1} and \cite{Schumacher W 04} for an excellent exposition
of quantum cellular automata and their rigorous definition and
classification).

The effect of the sequence given in Fig \ref{fig2} is a
transformation
\begin{eqnarray*}
    X_n &\rightarrow& \prod_{k=1}^{n} Z_k\\
    Z_n &\rightarrow& X_{n}X_{n+1}
\end{eqnarray*}
where $Z_{L+1}=\id_L$ if the system consists of $L$ spins and we use the
abbreviation $X_k\equiv \id^{\otimes (k-1)}\otimes X\otimes \id^{\otimes
(L-k)}$.
Connoisseurs will recognize this transformation as a special 
example of a duality transformation \cite{Kogut 79} which can map
for example the strong coupling regime of a Hamiltonian onto 
the weak coupling regime. Generalizations of such duality 
transformations for Ising models with multi-qubit interactions 
\cite{Turban 82} 
may be constructed easily by similar sequences of CNOT gates. Let 
us now apply this transformation to the Ising model in transverse 
magnetic field (whose strength we have set to $B=1$ for convenience) 
whose Hamiltonian, with open boundary conditions, is given by
\begin{equation}
    H(J) = \sum_{k=1}^{L-1} J X_kX_{k+1} + \sum_{k=1}^{L}
    Z_k.
\end{equation}
Then the above transformation leads to
\begin{eqnarray*}
    H_T(J) &=& TH(J)T^{\dagger} = X_L -JZ_1 + \sum_{k=1}^{L} J Z_{k} +
    \sum_{k=1}^{L-1} X_kX_{k+1}\\
    &=& X_L - JZ_1 + J H(J^{-1}).
\end{eqnarray*}
Evidently, the two Hamilton operators are not identical but it is
reasonable to expect that in the limit $L\rightarrow\infty$ the
two terms $X_L$ and $JZ_1$ may be neglected so that we can state
\begin{equation}
    H_T(J) \cong J H(J^{-1})\, .
    \label{scaling}
\end{equation}
Because the operators on both sides obey the same algebra this may
then be viewed as a statement about the symmetry of the Hamilton
operator itself so that we expect for the energy eigenvalues
$E(J)=J E(J^{-1})$ \cite{Kogut 79}. With this relation we may
conclude that if the spectral gap of the Hamiltonian vanishes for
one set of values $J$ then it will also vanish for $J^{-1}$
\cite{Fradkin S 78}. Thus the assumption of a single critical
point would then force the conclusion that it must be found at
$J=1$. This turns out to be correct in this case and indeed, near 
the critical point, the gap above the ground state for example 
is given by $\Delta E=2|1-J^{-1}|$ thus satisfying eq. 
(\ref{scaling}). The validity of these arguments are however 
perhaps not quite as nontrivial as one may expect as it is not 
per-se clear that a small perturbation at the boundaries leaves 
the spectrum essentially unaffected. Such a property requires 
proof (in fact and perhaps not too surprising
this is not correct in the extreme cases $J=0$ and $J=\infty$)
even though one can expect it to hold in most physically
reasonable cases.

Now we move beyond self-duality and link the cluster Hamiltonian
\cite{Pachos P 04,Kay LPPRR 05} to the an-isotropic XY model. The
cluster Hamiltonian, exhibiting some interesting critical
behaviour \cite{Pachos P 04} and strong finite size effects
\cite{Kay LPPRR 05}, is defined as
\begin{equation}
    H = \sum_{k=2}^{L-1} J X_{k-1}Z_kX_{k+1} + \sum_{k=1}^{L}
    B Z_k\, .
    \label{cluster}
\end{equation}
Firstly, employing a sequence of controlled phase gates between
any neighboring pair reveals that the model is self-dual in the
sense of the Ising model discussed above. Employing again the gate
sequence in Fig. \ref{fig2} we find that the Hamiltonian eq.
(\ref{cluster}) is mapped onto
\begin{equation}
    H_T = BX_L - \sum_{k=2}^{L-1} J Y_kY_{k+1} + \sum_{k=1}^{L-1}
    B X_kX_{k+1}\, .
    \label{clusterdual}
\end{equation}
Again, in the limit $L\rightarrow\infty$ we expect the
correspondence between the cluster Hamiltonian and the anisotropic
XY model to become exact and thus relating the critical behaviour
of the two models. Note however that the $BX_L$ term cannot be
neglected in the extreme limit $J=0$. Indeed, for $J=0$ the
Hamiltonian in eq. (\ref{cluster}) has a unique ground state while
eq. (\ref{clusterdual}) without the term $BX_L$ term would be
two-fold degenerate.

It is also straightforward to see that by the same transformation
the Hamiltonian
\begin{equation}
    H = \sum_{k=2}^{L-1} J_{1} X_{k-1}Z_kX_{k+1} +
    \sum_{k=1}^{L-1} J_{2} X_{k}X_{k+1} + \sum_{k=1}^{L}
    B Z_k \, .
    \label{cluster2}
\end{equation}
is mapped to
\begin{equation}
    H_T = BX_L - \sum_{k=1}^{L-2} J_{1} Y_kY_{k+1} + \sum_{k=1}^{L-1}
    B X_kX_{k+1} + \sum_{k=2}^{L} J_{2} Z_k\, .
\end{equation}
which is the $XY$-model in a transverse field whose critical
behaviour is well known. 
\ignore{Considering again eq. (\ref{cluster2})
and applying first a set of Hadamard transformation followed by
controlled phase gates on all neighboring pairs, again followed by
Hadamard transformations yields, in the asymptotic limit,
\begin{equation}
    H = \sum_{k} [J_1 Z_k + B X_{k-1}Z_kX_{k+1} +
    J_{2} X_{k}X_{k+1} ] \, .
    \label{cluster2dual}
\end{equation}
one discovers that this Hamiltonian is again self-dual.} Employing
operations such as those in figures \ref{fig2} and \ref{fig1} one
may obtain a large number of relationships between Hamilton
operators that become exact in the asymptotic limit. There are
various possible directions in which to extend such an approach.
One may for example consider the transformations that emerge from 
the Trotter decomposition of the time-evolution of Hamiltonians $H$ 
that one has decomposed into two parts $H_1$ and $H_2$ such that 
$H=H_1+H_2$ and $[H_1,H_2]=0$ \cite{Osborne E 06}.
Strictly, speaking it is not necessary to consider only unitary
operations. Needless to say that in this case the spectra of the
Hamiltonians are not connected in a very transparent way.
The next example demonstrates that such an approach
may nevertheless be useful. The following example will also serve
to demonstrate that in some cases the above approach actually
allows in a simple way to obtain the exact solution of certain
Hamilton operators. An example for that, which also serves to
motivate the definition of generalized stabilizer states, is

\noindent {\bf Lemma I --} {\em The ground state $|\Psi\rangle$ of
the translation invariant Hamiltonian
\begin{equation}
    H = \sum_{k=1}^{N} \left[-JZ_{k-1}X_kZ_{k+1} + B Z_k\right]
    \label{zxz}
\end{equation}
with periodic boundary conditions is uniquely determined by the
$N$ eigenvalue equations of non-Hermitean operators
\begin{equation}
    Z_{k-1} \left(\begin{array}{cc} 0 & \lambda \\ \lambda^{-1} &
    0\end{array}\right)_kZ_{k+1} |\Psi\rangle = |\Psi\rangle
    \label{ev}
\end{equation}
for all $i$ and
\begin{equation}
    \lambda = -\frac{B}{J} + \frac{J}{|J|}\sqrt{\left(\frac{B}{J}\right)^2 + 1}.
\end{equation}
}
\noindent {\bf Proof:} The strategy in the following will be to
map the Hamiltonian eq. (\ref{zxz}) to a sum of single particle
Hamiltonians that can be solved trivially. To this end, consider
the operator
\begin{equation}
    T = \prod_{j=1}^{N}
    \left(\begin{array}{cc} \lambda^{1/2} & 0  \\ 0 & \lambda^{-1/2}
    \end{array}\right)_{\!\!\! j} \,\,\prod_{k=1}^{N} U_{k,k+1}
\end{equation}
where we define the controlled phase $U_{k,k+1}$ acting on the
qubits $k$ and $k+1$ as
\begin{equation}
    U_{k,k+1} = \left(\begin{array}{cccc}
    1 & 0 & 0 & 0 \\
    0 & 1 & 0 & 0 \\
    0 & 0 & 1 & 0 \\
    0 & 0 & 0 & -1
    \end{array}\right).
\end{equation}
From the observation that the well-known cluster-state
$|Cluster\rangle =\prod_{k=1}^{N} U_{k,k+1} |+\rangle^{\otimes N}$
is uniquely determined as the state satisfying for all $i$ the
eigenvalue equations $Z_{i-1} X_i Z_{i+1}|Cluster\rangle =
|Cluster\rangle$ \cite{Briegel R 01} we immediately find that
\begin{equation}
    |\Psi\rangle = T |+\rangle^{\otimes N}
\end{equation}
satisfies the eigenvalue equations (\ref{ev}). The transformed
Hamilton operator $H_T = T^{-1}HT$ is given by
\begin{equation}
    H_T = \sum_{k=1}^{N} \left(\begin{array}{cc}
    B & -J\lambda^{-1} \\ -J\lambda &
    -B\end{array}\right)_{\!\! k}.
\end{equation}
Now we need to determine $\lambda$ such that the ground state of
$H_T$ is given by $\prod_{j=1}^{N}|+\rangle_j$. This is easily
found to be
\begin{equation}
    \lambda = -\frac{B}{J} + \frac{J}{|J|}\sqrt{\left(\frac{B}{J}\right)^2 + 1}
\end{equation}
and
\begin{equation}
    E_0 = -N \sqrt{B^2 + J^2}.
\end{equation}
The observation
\begin{equation}
    H|\Psi\rangle=E_0|\Psi\rangle \Leftrightarrow H_T|+\rangle^{\otimes N}=E_0
    |+\rangle^{\otimes N}.
\end{equation}
then completes the proof.\\

{\em Remark I --} We could have also applied a different
transformation $R$ which applies the local and non-local
operations in reverse order to that in $T$. Indeed
\begin{equation}
    {\tilde R} =  \prod_{k=1}^{N} U_{k,k+1}
    \prod_{j=1}^{N} U_j
\end{equation}
where $U_j$ is the unitary transformation that makes
\begin{equation}
    U_j^{\dagger} \left(\begin{array}{cc} B & -J  \\ -J & -B
    \end{array}\right)_j U_j
\end{equation}
diagonal and smaller eigenvalue being the vector
$|1\rangle=\left(0 \atop 1\right)$. The transformation on the
Hamilton operator is presented pictorially in Fig. \ref{fig1}.
Thus the state $|\Psi\rangle$ of eq. (\ref{ev}) may also be
written as $|\Psi\rangle = R|1\rangle$. It is noteworthy that by
interchanging the order in which the single particle and the
two-particle operations are applied changes the single particle
operator from a non-unitary to unitary.
This approach also allows immediately for the determination of the
entire eigenvalue spectrum. The energy separation between
eigenstates is simply given by the splitting between the
eigenvalues of $[B  -J; -J  -B]$. Thus it becomes evident that the
Hamilton operator does not possess a quantum critical point.

\noindent {\em Remark II --} For $J=0$ the ground state is a
product state, while for $B=0$ the ground state is a cluster
state. The transition between those two regimes is continuous and
thus we observe that the states $|\Psi\rangle$ can realize local
entropies ranging continuously in the interval $[0,1]$. This is in
contrast to standard stabilizer states whose local entropies are
quantized in integer units which in itself is enough to see that
standard stabilizer states will be ground states to a very limited
set of Hamiltonians  (see \cite{vandenNest LDB 06} for a much more
detailed discussion).\\

The simple example above suggests that the extension of the
stabilizer formalism to non-Hermitean operators may be useful. To
remind the reader, a 'stabilizer operator' for an N-particle state
is a tensor product of $N$ operators picked from the set
$\{X,Y,Z,\id\}$. A set $G=\{g_1,\ldots,g_N\}$ of $N$ mutually
commuting and independent stabilizer operators \cite{Independent}
is then called a 'generator set' uniquely identifying a state
$|\Psi\rangle$ that satisfies $g_k|\Psi\rangle=|\Psi\rangle$ for
all $k=1,\ldots,N$. Now we generalize these notions to the
definition of {\em generalized stabilizer states}.

\noindent {\bf Definition I (Generalized stabilizer states)} --
{\em For $N$ qubits, a generalized stabilizer state $|\Psi\rangle$
is the unique eigenstate to eigenvalue $1$ of the $N$ mutually
commuting and independent 'generalized stabilizer operators'
$\{g_1,\ldots,g_N\}$ where each $g_k$ is an N-fold tensor product
of arbitrary, possibly non-Hermitean, linear operators.}

From this definition we can immediately draw some straightforward
conclusions on some general classes of states that admit such a
generalized stabilizer state description.

\noindent {\em Remark III --} Note that any pure two-qubit state
can be written as a generalized stabilizer state. This follows
directly from the fact that each pure two qubit state can be
obtained from a maximally entangled singlet state by the local
application of linear operators. For three qubits all members of
the GHZ class are generalized stabilizer states while the members
of the W-class may be approximated arbitrarily well by generalized
stabilizer states (this follows from the well-known classification
of pure three-qubit states under the action of local linear maps
\cite{Dur VC 00}.) Some interesting results have been obtained recently 
in \cite{Hayashi MMOV 05,Markham MV 06,Dahlsten P 06} concerning 
the entanglement
content of certain stabilizer states and it would be interesting
to see how these results may be generalized to the setting of
generalized stabilizer states.

Note that one may extend the class of states which admit a
description also in another direction, namely by grouping together
neighbouring qubits. For $kN$ qubits, a generalized stabilizer state
$|\Psi\rangle$ is the unique eigenstate to eigenvalue $1$ of the
$N$ mutually commuting and independent 'generalized stabilizer
operators' $\{g_1,\ldots,g_{N}\}$ where each $g_k$ is an N-fold
tensor product of arbitrary, possibly non-Hermitean, linear
operators acting on $k$ qubits. It is evident that this will, 
for $k=3$ for example, describe all
possible three-qubit states but of course this comes at the expense of an
exponential increase in complexity of description. In some sense
the above concepts of generalized stabilizer states are included
in other descriptions of quantum states. In fact, for qubits
deformed weighted graph states as they have been introduced in
\cite{Anders PDVB 06} (see also \cite{Calsamiglia HDB 05,Dur HHLPB
05} for the concept of weighted graph states) incorporate the
generalized stabilizer states according to Definition I. However,
in this picture the description is again on the level wave-functions
\cite{footnote2}, i.e. the Schr{\"o}dinger picture, while the 
present approach, via
eigenvalue equations, is situated in the Heisenberg picture.
The weighted graph state picture of \cite{Anders PDVB 06,
Calsamiglia HDB 05,Dur HHLPB 05} is probably too general to allow
for a detailed quantification of multi-particle entanglement
while the generalized stabilizer states may well admit more 
detailed results for admittedly a less general class of states.
This will be the subject of a future publication.

As a final remark it should be noted that by definition,
generalized stabilizer states possess a unique characterization of
the quantum state of an n-qubit system employing only resources
that are polynomial in $n$. This alone, however, is not sufficient
for applications. It is also important to be able to derive
relevant physical quantities directly from the stabilizer
formalism. Indeed, having first to deduce the state explicitly and
then computing the property from the state would generally involve
an undesirable exponential overhead in resources. While one can
expect a direct approach to be possible in principle, it is
evident that detailed and explicit presentations of algorithms to
achieve these tasks in a systematic way and whose convergence is
proven are of interest. For standard stabilizer states such a
program has been carried out and detailed algorithms have been
provided \cite{Audenaert P 05}. There, normal forms in the context
of bi-partite entanglement have been found together with
algorithms with proven convergence to obtain these. These tools
may be transferred directly to generalized stabilizer states
whenever they are obtained from standard stabilizer states by the 
local application of linear operators.

\section{Conclusions}
In this work we have studied duality relations between Hamilton
operators from the viewpoint of sequences of quantum operations.
The close relationship of these transformations to duality
transformations in Hamilton operators has been noted. The same
approach has then been used to derive an exact solution for a
many-body Hamiltonian which has led to the characterization of the
ground state via eigenvalue equations using non-Hermitean
operators. This motivated the definition of a generalization of
the concept of stabilizer states to incorporate non-Hermitean
operators. The properties of such states and the potential offered
by this approach still remain to be explored.\\

{\bf Acknowledgements --} This work is part of the EPSRC QIP-IRC
(GR/S82176/0) and was supported by the European Integrated Project
QAP. MBP holds a Royal Society Wolfson Research Merit Award. The
author thanks Jiannis Pachos for pointing him to the concept of
duality when working on \cite{Pachos P 04} in 2003 and Oscar 
Dahlsten for comments.


\begin{thebibliography}{99}
%
\bibitem{Plenio V 07} M.B. Plenio and S. Virmani, Quant. Inf.
Comp. {\bf 7}, 1 (2007).
%
\bibitem{blockentropy} K. Audenaert, J. Eisert, M.B. Plenio,
R.F. Werner, Phys. Rev. A {\bf 66}, 042327 (2002); J.I. Latorre,
E. Rico and G. Vidal, Quant. Inf. Comp. {\bf 4}, 48 (2004); J.P.
Keating and F. Mezzadri, Commun. Math. Phys. {\bf 252}, 543
(2004).
%
\bibitem{Plenio CDE 05}  M.B. Plenio, M. Cramer, J.
Dreissig and J. Eisert, Phys. Rev. Lett. {\bf 94}, 060503 (2005);
M. Cramer, J. Eisert, M.B. Plenio and J. Dreissig, Phys. Rev. A
{\bf 73}, 012309 (2006); M. Cramer, J. Eisert and M.B. Plenio,
E-print arxiv quant-ph/0611264.
%
\bibitem{Fannes NW 92} M. Fannes, B. Naechtergaele and R.F.
Werner, Comm. Math. Phys. {\bf 144}, 443 (1992).
%
\bibitem{Verstraete C 04} F. Verstraete and J.I. Cirac, E-print
arxiv quant-ph/0407066.
%
\bibitem{Anders PDVB 06} S. Anders, M.B. Plenio, W. D{\"u}r, F.
Verstraete and H.-J. Briegel, Phys. Rev. Lett. {\bf 97}, 107206
(2006).
%
\bibitem{Dur HHLPB 05} W. D{\"u}r, L. Hartmann, M. Hein, M. Lewenstein
and H.-J. Briegel, Phys. Rev. Lett. {\bf 94}, 097203 (2005); M.
Hein, W. D{\"u}r, J. Eisert, R. Raussendorf, M. Van den Nest, and
H.-J. Briegel, Proceedings of the International School of Physics
"Enrico Fermi" on "Quantum Computers, Algorithms and Chaos",
Varenna, Italy, July 2005, E-print arxiv quant-ph/0602096.
%
\bibitem{Schumacher W 04} B. Schumacher and R.F. Werner, E-print
arxiv quant-ph/0405174.
%
\bibitem{Kogut 79} J. B. Kogut, Rev. Mod. Phys. {\bf 51}, 659
(1979).
%
\bibitem{Turban 82} L. Turban, J. Phys. C {\bf 15}, L65 (1982).
%
\bibitem{Fradkin S 78} E. Fradkin and L. Susskind, Phys. Rev. D {\bf
17} 2637 (1978).
%
\bibitem{Pachos P 04} J.K. Pachos and M.B. Plenio, Phys. Rev.
Lett. {\bf 93}, 056402 (2004).
%
\bibitem{Kay LPPRR 05}
A. Key, D.K.K. Lee, J.K. Pachos, M.B. Plenio, M. E. Reuter, and E.
Rico, Optics and Spectroscopy {\bf 99}, 339 (2005).
%
\bibitem{Osborne E 06} J. Eisert and T.J. Osborne, Phys. Rev. Lett. 
{\bf 97}, 150404 (2006)
%
\bibitem{Briegel R 01} H.-J. Briegel and R. Raussendorf,
Phys. Rev. Lett. {\bf 86}, 910 (2001).
%
\bibitem{Independent} A set of stabilizer operators is called
independent if $\prod_{i=1}^N g_i^{s_i}=\id$ exactly if all $s_i$
are even.
%
\bibitem{Dur VC 00} W. D{\"u}r, G. Vidal and J.I. Cirac, Phys.
Rev. A {\bf 62}, 062314 (2000).
%
\bibitem{Audenaert P 05} K.M.R. Audenaert and M.B. Plenio,
New J. Phys. {\bf 7}, 170 (2005).
%
\bibitem{Hayashi MMOV 05} M. Hayashi, D. Markham, M. Murao, M.
Owari and S. Virmani, Phys. Rev. Lett. {\bf 96}, 040501 (2006)
%
\bibitem{Markham MV 06} D. Markham, A. Miyake and S. Virmani,
E-print arxiv quant-ph/0609102
%
\bibitem{Dahlsten P 06} O.C.O. Dahlsten and M.B. Plenio, Quant. Inf. 
Comp. {\bf 6}, 527 (2006).
%
\bibitem{Calsamiglia HDB 05} J. Calsamiglia, L. Hartmann, W.
D{\"u}r and H.-J. Briegel, Phys. Rev. Lett. {\bf 95}, 180502
(2005).
%
\bibitem{footnote2} Note however that the numerical algorithms
presented in \cite{Anders PDVB 06} make use of a mixed picture
partially in the Schr{\"o}dinger and partially in the Heisenberg
picture.
%
\bibitem{vandenNest LDB 06} M. Van den Nest, K. Luttmer, W.
D{\"u}r and H.-J. Briegel, E-print arxiv quant-ph/0612186.
%
\end{thebibliography}
\end{document}